\documentstyle[psfig,epsfig,12 pt]{article}

\def\g{\gamma} 
\def\a{\alpha}

\def\d{\delta}

 \def\L{\Lambda}
\def\m{\mu}
\def\n{\nu}

\def\mn{{\mu\nu}}
\def\be{\begin{equation}}
\def\ee{\end{equation}}

\begin{document}

\begin{flushright} BRX TH-497
\end{flushright}

\begin{center} {\Large\bf New Properties of Matter in (A)dS\\ and
their Consequences}

S. Deser\footnote{Invited talk at Journe\'{e}s Relativistes, Dublin, 2001.
Dedicated to the memory of Lochlain O'Raifeartaigh. \\
\hspace*{.2in} \tt deser@brandeis.edu} \\
Department of Physics, Brandeis University\\ Waltham, MA 02454,
USA
\end{center}

\noindent{\bf Abstract}: I review briefly, primarily for
relativists, a series of recent results, obtained with A.\
Waldron, on the novel behavior of massive higher ($s>1$) spin
systems in constant curvature backgrounds.  We find that the
cosmological constant $\L$, together with the mass parameter,
define a ``phase plane" in which partially massless gauge
invariant lines separate allowed regions from forbidden,
non-unitary, ones.  These lines represent short multiplet systems,
with missing lower helicities, removed by novel local gauge
invariances, and (despite having $m \neq 0$) propagating on the
light cone.  In the limit of an infinite tower of these higher
spin bosons and fermions, unitarity requires $\L$ to vanish.

\vspace{.1in}

The kinematical effects of gravity on matter (as against the
well-known dynamical ones) have not received much attention in the
past, nor would one intuitively expect any major surprises there.
I will report here (primarily) on a series of very recent
investigations by A.\ Waldron and myself \cite{001} in which
``massive" higher spin ($s>1$) free bosons and fermions exhibit
unexpected, qualitative, differences from flat space in the
simplest curved backgrounds -- constant curvature (deSitter)
spaces -- denoted collectively by (A)dS to cover both
(negative)/positive cosmological constants $\L$.  Concepts
synonymous in Minkowski geometry, such as masslessness, light cone
propagation, maximal helicity modes only, and gauge invariance
become nondegenerate, giving rise to such exotic effects as null
propagation, partial (local) gauge invariances and shortened
helicity multiplets, all with $m \neq 0$.  Furthermore, entire
ranges of mass become forbidden by unitarity.  Perhaps most
dramatically, accomodation of towers of excitations of all
possible spins is only possible in the limit of vanishing $\L$,
thus providing a dramatic, if not quite yet physical, solution of
the cosmological problem.  All this happens because the dull flat
space mass line is here replaced by a plane, parametrized by the
dimensional duo ($m^2 , \L$).  I will also briefly mention the
relation of all this to the old problem of the $m^2 \rightarrow 0$
limit discontinuity in matter-matter interactions \cite{002} and
its Newtonian counterpart \cite{003}.

Because my space is very limited, and most of the results are now
available, I will only skim the highpoints just summarized.
Indeed, let me devote a good fraction of this space to a single
picture that summarizes many of these results.

\vspace{1.cm}
\begin{center}
\epsfig{file=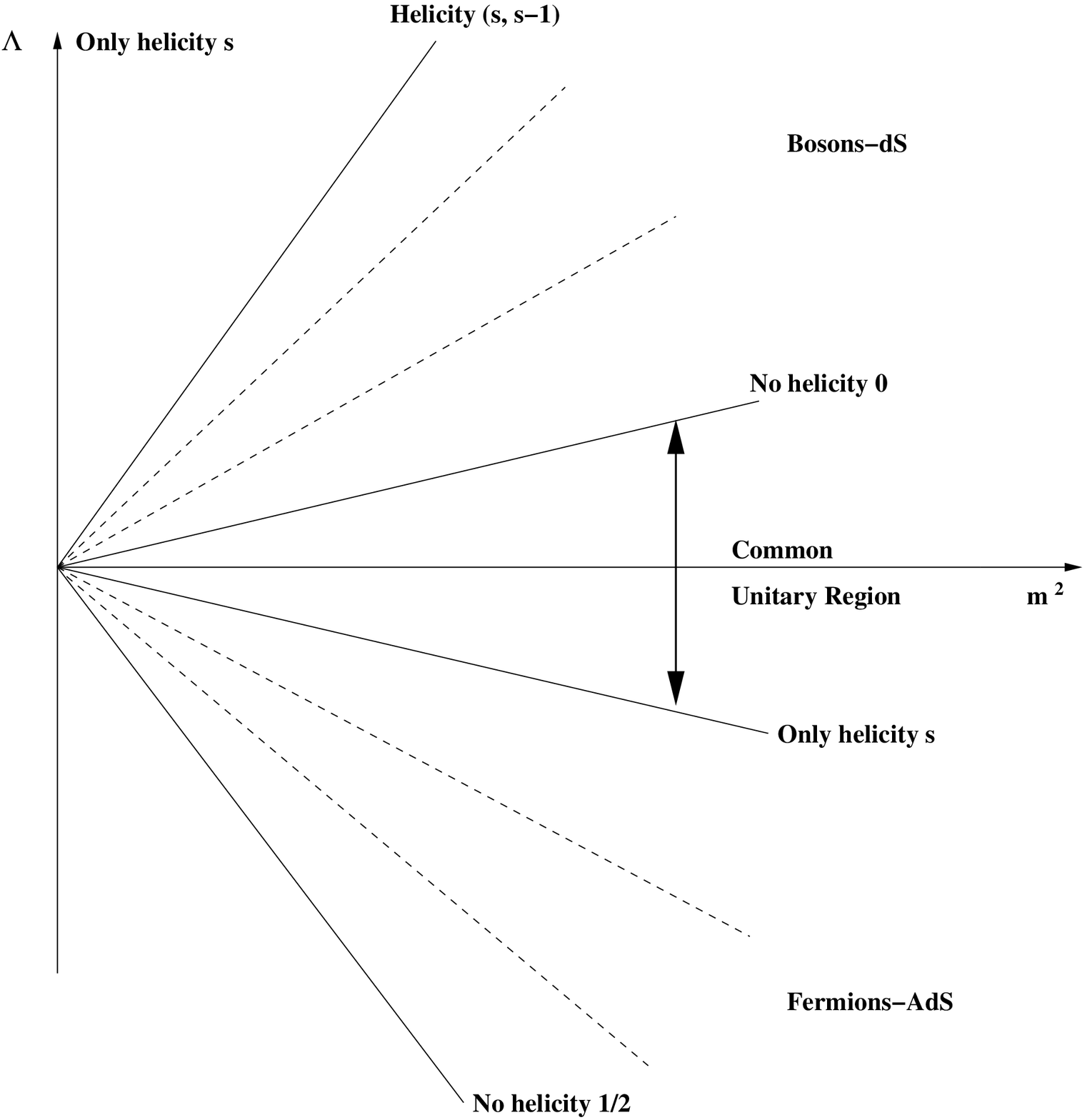, width= 15.cm}
\end{center}
\baselineskip=13pt \centerline{}

\noindent\footnotesize{{\bf Figure 1}:  The top/bottom halves of
the half-plane represent dS/AdS (and also bosons/fermions)
respectively. The $m^2=0$ vertical is the familiar massless
helicity $\pm s$ system, while the other lines in dS represent
truncated (bosonic) multiplets of partial gauge invariance: the
lowest has no helicity zero, the next no helicities $(0,\pm 1)$,
etc.  Apart from these discrete lines, bosonic unitarity is
preserved only in the region below the lowest line, namely that
including flat space (the horizontal) and all of AdS. In the AdS
sector, it is the topmost line that represents the pure gauge
helicity $\pm s$ fermion, while the whole region below it,
including the partially massless lines, is non-unitary. Thus, for
fermions, only the region above the top line, including the flat
space horizontal and all of dS, is allowed.  Hence the overlap
between permitted regions straddles the $\L = 0$ horizontal and
shrinks down to it as the spins in the tower of spinning particles
grow; only $\L = 0$ is allowed for generic ($m^2$ not growing as
$s^2$) infinite towers.}

\newpage

\normalsize
 In the remaining pages, I motivate these
results in terms of a typical -- and most familiar to relativists
-- system, namely spin 2, followed by an even briefer excursion
into the fermion case. Consider then, the massive Pauli--Fierz
system in a cosmological background, whose equations are
 \be
 {\cal G}_\mn (\phi ) + m^2 (\phi_\mn - \bar{g}_\mn \phi^\a_\a ) =
 0 \; .
 \ee
Here all metrics, indices and covariant derivatives are those of
the cosmological background, and $\phi_\mn$ the spin 2 field. Thus
${\cal G}_\mn$ is just the linearized deviation about (A)dS of the
Einstein metric $g_\mn = \bar{g}_\mn + \phi_\mn$; the minor
$([D_\m , D_\n ] \sim \L )$ ambiguity in the order of the two
covariant derivatives in ${\cal G}_\mn$ is determined by requiring
the order in which the Bianchi identity $\bar{D}_\m {\cal G}^\mn
(\phi) \equiv 0$ holds.

 Now let us proceed as in flat space and determine the
 consequences of taking successive divergences eq. (1).  It
 follows directly that (of course)
 \be
\bar{D}_\m \, {\cal G}^\mn = -m^2 \, \bar{D}_\m \, (\phi^\mn -
\bar{g}^\mn \phi^\a_\a )\; ,
 \ee
 but then comes the surprise
 \be
\bar{D}_\m \bar{D}_\n \, {\cal G}^\mn + \frac {1}{2} \, m^2 {\cal
G}^\m_\m =  3/2 m^2 \, (m^2 - 2/3 \, \L ) \phi^\a_\a \; .
 \ee
 [For higher spins, further divergences are possible and generate
 additional terms of the form $(m^2 - \a_i \L )$ on the right
 side.]  The first step, (2), is super-familiar: If $m^2 =
 0$, the theory possesses the invariance $\d\phi_\mn = \bar{D}_\m
 \phi_\n + \bar{D}_\n \phi_\m$ and hence has the usual two degrees
 of freedom (DoF) corresponding to helicity $\pm$2 (a valid
 concept in (A)dS).  If $m^2 \neq 0$ we expect the loss of gauge
 invariance to permit the 5 helicity states $(\pm 2, \; \pm 1, \;
 0)$.  But now look at (3): while it teaches us nothing further
 for $m^2 \equiv 0$, there appears a new zero, at $m^2 = 2\L /3$.
 This is the first appearance of partial masslessness/gauge
 invariance.  Partial masslessness because in fact this $m^2\neq
 0$ system nevertheless has null propagation just like the $m^2 =
 0$ one (remember that (A)dS, being conformally flat, shares the
 usual Minkowski light cone).  Gauge invariance because it is
 clear from (3) that the action leading to (1), namely $\frac{1}{2}\int
 \phi_\mn [{\cal G}_\mn (\phi ) + m^2 (\phi_\mn - \bar{g}_\mn \phi^\a_\a )]$
 is invariant under the reduced,
 but still local gauge invariance
  \be
\d \phi_\mn = (\bar{D}_\m \bar{D}_\n + \bar{D}_\n \bar{D}_\m + 2\L
/ 3 \bar{g}_\mn ) \xi (x) \; ,
 \ee
since $\d I [\phi ] = \int \d \phi_{\mn} [{\cal G}^{\mn} + m^2
(\phi^\mn - \bar{g}^\mn \phi ) ]$ .  But a gauge invariance
removes a degree of freedom, in our case that of helicity zero.
This system was actually discovered earlier \cite{004}, as was its
DoF content \cite{005}. The actual mechanism of this amputation in
a Hamiltonian analysis of the system makes for a very amusing (if
messy) calculation, but the real surprise here is that, as one
varies $m^2$ in the $(m^2 , \L )$ plane in a given ($\L$ fixed)
geometry, this excitation turns from being a normal one to
vanishing, then reemerging as a ghost -- thereby generating a
unitarily forbidden region as shown in the top (dS) part of Fig.
1.  Let me just write the relevant part of the helicity zero
Hamiltonian to show how this works:
\newpage

\be
H \approx \textstyle{\frac{1}{2}} [\n^{-2} p^2 + \n^2 (\nabla
q)^2]\; .
 \ee
Clearly, if $\n^2 \equiv (m^2 - 2 \L / 3)$ is positive (as it is
for $\L = 0$) an obvious rescaling $(p,q)\rightarrow (\n p,
\n^{-1} q)$ will give a normal action. However, in the region
$\n^2 < 0$, the other side of the (partial) gauge line, this can
only be accomplished at the price of imaginary variables or
negative energy.  Thus, the $\n^2 = 0$ line (where it is easy to
see from a previous step that the whole helicity 0 action
vanishes) serves as a divider between the good and forbidden
regions of the $(m^2 , \L )$ plane.

While the above interesting behavior for bosons takes place in dS
$(\L > 0)$, fermions prefer AdS for reasons we do not yet
understand. The simplest case is $s=3/2$, which actually does not
display the novel behavior; that starts with the tensor-spinor
$\psi_\mn$ of $s=5/2$.  Neverthless, it illustrates another deep
fact (known from the birth of supergravity), that when $\L \neq
0$, gauge invariance for $s=3/2$ is NOT displayed at $m$=0 (as it
is for bosons) but rather at a finite mass parameter, and only in
AdS $(\L < 0)$. Recall that the field equation reads
\be
 R^\m \equiv \g^{\mn\a} {\cal D}_\n \psi_\a = 0\; , \;\;\; {\cal
 D}_\n \equiv D_\n + \textstyle{\frac{1}{2}} \; \g_\m \; m
 \ee
and that the gauge invariant -- hence truly ``massless" -- system
occurs at ${\cal D}_\m R^\m = 0$, namely for $m^2
+\textstyle{\frac{1}{3}}\;  \L = 0$, where $[{\cal D}_\m , {\cal
D}_\n ] \psi_\a \equiv 0$.  The helicity $\pm$1/2 component of the
system mimics the behavior of helicity 0 for spin 2, but in the
``inverted", AdS, sector:  Above the gauge line it is present and
unitary, below it it is non-unitary and is of course absent
entirely at $m^2 + \frac{1}{3} \: \L = 0$.  Put another way, the
additional difference is that, unlike bosons, where $s=1$ has only
one index and so no novel behavior, already at $s=3/2$ (and
beyond) ``true" masslessness, {\it i.e.}, complete (not partial)
gauge invariance occurs not at $m=0$, but rather, generically, at
$m^2 = - |\a |\L$.  This is due to the spinor part of the
spinor-vector (or -tensor) field, and was already discovered by
Dirac very long ago for $s=1/2$. Essentially, when one squares the
Dirac equation, $(\g \cdot D + m) (\g \cdot D - m) \psi = 0$,
there is a residue proportional to ($m^2 + |\a |\L$) due to the
famous identity $(\g \cdot D)^2 \equiv D^2  + R/4$.  That one is
totally separate from and just adds to the partial mass effect
from the world indices.

To complete our navigation of Fig. 1, we note first that partial
gauge systems defined by the transition lines correspond to
truncated multiplets from which one or more of the lowest
helicities are missing, but that in each case, the remaining
(higher than 0 or $\pm$1/2) helicities all propagate on-cone.
Secondly, the gauge lines for generic spin are governed
respectively by $m^2_B \sim \frac{1}{3} \: \L s(s-1)$ and $m^2_F
\sim -\frac{1}{3} \: \L (s-\frac{1}{2})^2$.  Hence if one
considers towers of rising spin particles of both statistics, and
if their masses are not tuned to rise with spin (or at least not
as fast as $s^2$) then since the allowed, unitary, region common
to bosons and fermions spans a cone around $\L = 0$, our infinite
tower (such as found in zero slope string expansions) is only
permitted at $\L\rightarrow 0$ in the limit.  This mechanism,
while simple and appealing, is not necessarily robust under more
physical circumstances, such as presence of dynamical gravity and
of other interactions.  In this connection, we mention that
consistent coupling of (even massive) higher spin fields to
gravity, is hard to achieve beyond supergravity.  This and the
additional problems for charged as well as gravitating systems has
recently been investigated systematically in the last paper of
\cite{001}.

\newpage

Finally, we turn to the question that triggered some of our
studies, namely the so-called vDVZ discontinuities of linearized
massive gravity and spin 3/2 interacting with -- necessarily
prescribed -- conserved matter sources \cite{002}.  Spin 2, in
flat space, for example, when coupled to a conserved external
stress tensor differs from its spin 1 counterpart in that the
helicity 0 mode fails to decouple from the source in the massless
limit, with the result that there is a large finite difference in
the prediction of light bending in the $m^2 \rightarrow 0$ and
$m^2 \equiv 0$ cases once they are both fixed to give the proper
Newtonian gravity limit, {\it i.e.}, the effective coupling differ
in the ratio of $t^\m_\m \, T^\n_\n$ to $t_\mn \, T^\mn$ couplings
between two sources, being 1/3 vs 1/2 respectively. When $\L \neq
0$, however, these ratios depend on both $(m^2 , \L )$ parameters
and can have almost any value depending on the path taken to
(0,0).  It is also instructive \cite{003} to discuss what happens
in the Newtonian $(c\rightarrow \infty )$ limit.  Here matter
essentially consists of only $T^0_0$ and $T^\m_\m \sim T^0_0 \neq
0$; there should be no discontinuity since there is no light whose
bending is to be explained.  Indeed, one obtains a result that
correctly reflects this physical expectation, but only after
realizing that apparent wrong -- repulsive or vanishing or
infinite Newtonian couplings are excluded either by the
nonunitarity of the offending ``gravitons", or by the truncated
multiplet case which, by the invariance (4), can only couple to
traceless conserved $T^\mn$ sources.

I thank A.\ Waldron and B.\ Tekin for stimulating collaborations.
This work was supported by NSF Grant PHY99-73935.


\begin{thebibliography}{99}
  \bibitem{001}
          S.\ Deser and A.\ Waldron, Phys.\ Rev.\ Lett.\ {\bf 87}, 031601
  (2001); S.\ Deser and A.\ Waldron, Nucl.\ Phys.\ {\bf B607}, 577--604
  (2001); S.\ Deser and A.\ Waldron, Phys.\ Lett.\ {\bf B508}, 347--353
  (2001); S.\ Deser and A.\ Waldron, Phys.\ Lett.\ {\bf B513}, 137--141
  (2001).
\bibitem{002}
 S.\ Deser and A.\ Waldron, Phys.\ Lett.\ Phys.Lett. {\bf B501}, 134--139 (2001)
for a history.
\bibitem{003}
S.\ Deser and B.\ Tekin, Class.\ Quant.\ Grav.\ {\bf 18}, L171
(2001).
\bibitem{004}
 S. Deser and R. Nepomechie, Phys. Lett. {\bf B132}, 321 (1983); Ann.
 Phys. {\bf 154}, 396 (1984).
\bibitem{005}
 A. Higuchi, Nucl. Phys. {\bf B282}, 397 (1987); {\it ibid.} {\bf
 325}, 745 (1989); J. Math. Phys. {\bf 28}, 1553 (1987).
\end{thebibliography}
\end{document}